# Development of a fracture capture simulator to quantify the instability evolution in porous medium

Ramesh Kannan Kandasami[1], Charalampos Konstantinou[2] and Giovanna Biscontin[3]

[1]Department of Civil Engineering, Indian Institute of Technology Madras, Chennai, India. Email: rameshkk@iitm.ac.in
[2]Department of Civil and Environmental Engineering, University of Cyprus, Nicosia, Cyprus. Email: ckonst06@ucy.ac.cy
[3]Department of Engineering, University of Cambridge, U.K. Email: gb479@cam.ac.uk

**ABSTRACT**

Understanding and controlling fracture propagation is one of the most challenging engineering problems, especially in the oil and gas sector, groundwater hydrology and geothermal energy applications. Predicting the fracture orientation while also possessing a non-linear material response becomes more complex when the medium is non-homogeneous and anisotropic. Fracturing behaviour in geological porous media that exhibit high leak-off potential is not clearly understood. In this context, a novel testing technique is used to simulate the ground conditions in the laboratory and study the instability characteristics of such geo-materials. The bespoke apparatus designed and developed in this research programme is capable of applying true anisotropic boundary stresses, injecting fluid at a predefined flow rate and viscosity while also imaging the instability/ fracture propagation in a porous medium such as sand and weak rock. Pressure profiles and the progression of fracture are recorded simultaneously during the fluid injection process into specimens subjected to different boundary stresses. The fracture propagation data are analysed, which provides information on the evolution of fracture morphology and expansion velocity during the injection event.

## 1. Introduction

Hydraulic fracturing is ubiquitous not only in oil well exploration but also in grouting industries (Morgenstern 1963), geothermal energy extraction (Zhang et al. 2020), subsurface hydrogen storage (Raza et al. 2022), $CO_2$ sequestration (Fu et al. 2017), coal bed methane extraction under frozen conditions (Fan et al. 2019), groundwater decontamination (Mahmoodi et al. 2020), hydraulic barriers (Konstantinou and Biscontin, 2022) and has given great impetus in indirect crustal stress measurements (Zoback et al. 1977). Extensive studies have been carried out to understand the generation and propagation of hydraulically induced fractures under complex boundary conditions, especially in competent rock formations (Guo et al. 1993, Rubin 1983, Durban and Papanastasiou 1997, Papanastasiou 1999 & 2000, Peterson et al. 2001). Mathematical models (Khristianovic and Zheltov 1955, Perkins and Kern 1961, Geertsma and Haafkens 1979, Desroches et al. 1994, Geertsma and De Klerk 1969, Nordgren 1972, Sneddon and Lowengrub 1969) based on the fundamental principles of elasticity, lubrication theory, basic assumptions on the material behaviour such as homogeneity, isotropy and impermeability were used to derive the fracture pressure and fracture dimensions. However, the fracture response in highly porous rock deviates substantially compared to a competent rock formation. Further, the existing experimental methods rely on indirect measurements of fracture characteristics which are generally unreliable. Simulating the in-situ stresses in the laboratory and studying the fracture morphology in porous media during a fluid injection event is essential to effectively understand the fracturing process in natural materials.

## 2. Overview of experimental techniques

The first laboratory study on hydraulic fracturing in rocks was carried out by Scott et al. (1953) on thick hollow cylinders of sandstone and shale. A specialized steel chamber was designed to receive the specimen while also applying the confining pressure. During fluid injection, the fracture pressure was measured, and comparisons were made with theoretical model predictions. In the ensuing research, Lamont and Jessen (1963) studied fracture propagation across existing discontinuities in rectangular block specimens using a true triaxial and biaxial setup. The fracture geometry was observed after post-test slicing of the specimen to quantify the influence of the existing plane of weakness on the direction of hydraulically induced fractures. Further, cubical and cylindrical hydro-stone specimens were hydraulically fractured in the laboratory (Haimson and Fairhurst, 1969) by applying true triaxial and biaxial stresses, respectively. In the case of a true triaxial system, the two horizontal stresses were applied using hydraulically powered flat jacks, while the vertical stress was applied using a compression testing loading frame. In the biaxial setup, the confining pressure was applied on the specimen placed inside a chamber filled with oil while the compression testing machine applied the axial stress. Two injection fluids (penetrating and non-penetrating fluid) were used for fracturing the porous rock specimens. The effect of the rate of fluid injection and injection hole diameter on the breakdown pressure was measured. Experimental studies were also performed by Rubin (1983), specifically in impermeable materials like PMMA (a transparent acrylic) instead of rock, under plane strain conditions in order to verify the predictions of fracture models. A specifically designed experimental setup was developed with a water-injection fluid interface (separation chamber), a hand pump to apply pressure to pre-fracture the model and a mechanical pump to apply the injection pressure at a constant flow rate - all three connected in series. Since PMMA was transparent, images were captured to record the fracture propagation. These experiments were performed without the application of confining pressure and on a non-representative material system.

In addition to the slew of hydro-fracturing experiments on quasi-brittle materials, Mori and Tamura (1987) accomplished a series of fracture tests on cohesive soils. A special triaxial fracturing apparatus was developed where the confining pressure and deviatoric stress were applied to a predetermined value and held constant while hydraulically fracturing the cohesive soils. They observed that the fracture initiates when the borehole pressure exceeds the sum of the confining pressure and shear strength of the soil. After fracturing the specimens, Rhodamine dye was injected through the borehole to investigate the fracture morphology. Murdoch (1993) studied the fracture patterns in clayey silt at different moisture content under low triaxial stresses. The fracturing cell consisted of inflatable neoprene rubber bladders capable of applying three-dimensional confining stresses on the rectangular block specimens. A starter slot was created in the specimen through which Rhodamine dyed glycerine was injected to induce fractures. After these tests, the specimens were cut to investigate the characteristics of fracture propagation. Studies on dense, purely frictional granular material (Konstantinou et al. 2021 and Huang et al. 2012) quantified the influence of injection fluid viscosity and flow rate on the cavity and finger formation by using a modified Hele-Shaw cell. Albeit without applying any boundary stresses, this setup was useful for capturing the grain displacement patterns in real time using a high-resolution camera. Analyzing the fracture morphology was also carried out using acoustic emission (Lockner and Byerlee, 1977) techniques in addition to post-test slicing and CT scanning to demarcate between tensile and shear failures.

Investigations were carried out to study the fracture pressure and fracture morphology post-fracturing (using CT scanning, post-test slicing) or indirect methods (acoustic techniques), while a real-time observation of fracture initiation and propagation under realistic boundary conditions, especially in poorly consolidated reservoirs, is still lacking. Due to the dearth in understanding of fracturing behaviour in porous geo-materials, the current research focuses on the design and development of a unique bespoke advanced laboratory experimental setup which is capable of applying true three-dimensional anisotropic boundary stresses on the specimen, injecting the fluid with predefined viscosity and flow rate while simultaneously imaging the fracture evolution characteristics. Experiments were carried out on unconsolidated sand and weak rock specimens using a fracturing fluid under different boundary conditions. The real-time visualization during fracturing experiments revealed the dependence of fracture propagation morphology on the boundary stresses and material properties.

## 3. Experimental setup
### 3.1 Fracture capture simulator

The current study presents a novel experimental set-up incorporating realistic stress confining conditions, in addition to real-time imaging, which is crucial in quantifying the infiltration front and characteristics of fracture evolution. The principal motivation for developing this device is to understand the fracture and leak-off characteristics of poorly consolidated and permeable formations. Thus, the potential specimens consist of highly permeable material systems ranging from clean sand to weak rock with an unconfined compressive strength between 0 and 2 MPa and having hydraulic conductivity between $10^{-2}$ to $10^{-6}$ m/s. The target boundary stresses should replicate in-situ conditions ranging from those encountered in shallow grouting (tens of kPa) to those observed during shallow depth hydraulic fracturing processes (tens of MPa). In order to keep the size of the device suitable for benchtop testing without cranes or special lifting equipment, the apparatus is designed to apply boundary stresses between 0 to 3 MPa.

The shape and size of the specimens are other key elements of the design. The apparatus should be capable of applying three-dimensional stresses independently. Most of the previous laboratory studies on hydraulic fracturing utilized cubical specimens, often referred to as "block specimens," to apply three principal stresses in mutually perpendicular directions using rigid platens. Problems arise with the application of independent stresses to flat surfaces, especially near the edges, particularly as the deformations develop. Prismatic pressure vessels or reaction frames are also much more difficult to build and less sturdy than cylindrical ones. Traditionally, cylindrical specimens have also been used for fracturing experiments (Downhole Simulation Cell - DSC, Simpson et al. 1989) but are limited in the single radially applied confining stress. The proposed device uses flexible, deformable membranes to apply different horizontal stresses on a cylindrical specimen. The specimen dimensions are selected to minimize boundary effects, provide sufficient room to capture fracture initiation and propagation and allow a reasonable zone of influence (i.e., the ratio of specimen diameter to injection hole diameter). After the initial calculations based on these key considerations, the specimen diameter is fixed at 150 mm and thickness at 40 mm.

In addition to the simulator, the ancillary components required for real-time visualization of fracture in porous geomaterials during fluid injection are shown in Fig. 1. The complete setup (Fig. 1) is composed of the fracture capture simulator placed in a hydraulic loading frame to apply the axial stress while the differential confining stress is applied using an air compressor. Further, the fluid is injected into the simulator using a high-pressure syringe pump. The compliance requirements between different components of the experimental setup are found to be satisfactory. The sensors measuring the pressure and displacement are connected to the data acquisition setup, which in turn is connected to the computer to record the data. A high-speed camera is placed at the base of the simulator to capture the leak-off and fracture propagation in real-time.

The fracture capture simulator (Fig. 2) itself consists of three essential components: the top cap, the base pedestal and the bladder holders.

*Top cap:* The three principal functions of the top cap are to apply uniform axial pressure onto the specimen, house drainage ports for proper drainage of the injected fluid and accommodate an injection port while also supporting the injection probe. The detailed representation of the top cap is given in the schematic (Fig. 3). The top cap is made of hardened aluminium and can handle a nominal axial load of 50 kN (equivalent to 2.8 MPa of pressure). Pockets are machined into the top cap, as shown in section YY (Fig. 2), to minimize its dead weight and avoid loading the specimen during the preparation process. Four drainage ports of 2.5 mm diameter are located diametrically opposite around the side circumference of the cap. The injection port starts on the outer boundary of the cap and traverses to the centre, as shown in cross-section XX (Fig. 3). Injection tubes of different heights and inner diameters can be used to deliver the fluid at a specific depth within the specimen. Since the top cap and Perspex window (present in base pedestal) are perfectly flat, it can be assumed that the axial pressure is uniformly applied on the specimen provided the specimen is prepared well and perfectly flat.

*Base pedestal and bottom slab:* The base pedestal consist of a Perspex window embedded in the base and also serves as the head for the columns while also providing drainage ports (Fig. 4). The Perspex window is transparent and acts as a base for placing the specimen. This window allows visually capturing fracture initiation and propagation during the injection process. Even though the Perspex window has the advantage of being clear and transparent, the low modulus and tensile strength of this material leads to limitations in the application of axial pressure to avoid excessive bending. In this study, the thickness of the Perspex sheet is fixed at 50 mm, resulting in a maximum allowable pressure of 3 MPa. The stress analysis carried out on this Perspex window showed a maximum differential deflection of 0.2 mm between the centre and the edge at a maximum uniformly applied pressure. There are four drainage holes symmetrically placed on the window in order to freely drain the injected fluid away from the viewing area, as presented in the top view of Fig. 4. The clear viewing space on the Perspex window has a diameter of 100 mm. The Perspex window is supported on an aluminium base which has four column heads. A rubber gasket is glued between the Perspex window and the aluminium base to prevent any leaks during the injection process. The four stainless-steel columns are attached to the bottom slab at the other end (Fig. 2). The space between the four columns hosts a high-speed camera pointing towards the Perspex window to image the fracturing process. The dimensions of the column and the clear distance between the columns are given in Fig. 2.

***Bladders and bladder holder:*** In order to apply differential confining pressure, four custom fabricated Ethylene Propylene Diene Monomer (EPDM) rubber bladders with nylon reinforcement were used. These curved rubber bladders are highly durable and can apply the desired pressure uniformly onto the circumference of the cylindrical specimen (Fig. 5). During a typical experiment, diametrically opposite rubber bladders are always kept at constant pressure while differential confining pressures can be applied by varying the pressures in the two adjacent bladders. Each rubber bladder occupies a quarter of a circular arc and can apply a pressure of up to 2 MPa when it is well contained. Bladder holders and packers contain the rubber bladders firmly in their position on three sides (top, bottom and back). The two bespoke holders, each housing two rubber bladders, act as a reaction element which is designed to withstand a pressure of 2 MPa when the bladders are pressuring the specimen. The dimensions of the rubber bladders, packers (see inset) and bladder holders are shown in section XX of Fig. 5. In addition to firmly connecting the bladder holders, the holders are secured to the base pedestal (Fig. 4) to contain the specimen firmly. The inlet valves are moulded to the rubber bladder during the manufacturing process (Fig. 5) to ensure a monolithic construction.

One of the important concerns in developing this test rig was to ensure the uniform application of boundary stresses. Even though the three sides (top, bottom and back) of the rubber bladders are fully confined by the holders to allow expansion only in one direction, it is extremely important to assess whether the applied pressure is transferred completely and uniformly onto the specimen. A tactile pressure sensor is wrapped around the specimen to verify the amount of stress transferred from the rubber bladders to the specimen. This thin sensor sheet consists of a matrix of "sensels" which measure the pressure in real-time. These tactile pressure sensor sheets are pre-calibrated before carrying out actual measurements. Fig. 6 shows a comparison between the applied pressure to inflate the bladders and the average pressure transmitted onto the specimen. The figure clearly shows that the difference in pressure levels (applied and measured) is within ±5−6%, and hence the boundary stress application is considered satisfactory. Another common issue observed while applying a three-dimensional stress field is the non-conformity of stresses near the boundaries or edges. There is a blind zone (about 0.5 mm width) between two adjacent bladders where the stresses shift. However, the advantage of using cylindrical specimens is that these blind spots are localized. By fixing the size of the specimen and injection hole diameter appropriately, i.e., if the ratio of specimen diameter to the injection hole diameter is greater than 20, the application of the boundary stresses will not affect the fracture characteristics.

**3.2 Injection system**

A positive displacement pump is used in this study to generate precise linear flow (without any pulsations) and avoid secondary effects on fracture initiation and propagation. An electro-mechanically operated syringe pump (shown in Fig. 1), capable of handling fluids of different viscosity and drilling mud concentration, is selected based on the maximum flow rate, injection pressure, and total volume required for fracturing the material systems envisioned in this study. A syringe pump from GDS instruments with a maximum flow rate of 30 mL/min, 3 MPa pressure and 180 mL volume is used for testing loose sand specimens, while a Teledyne ISCO pump with a maximum flow rate of 200 mL/min, 30 MPa pressure and 500 mL volume is used for testing weak rock specimens. The pump is connected to the injection port located on the top cap of the fracture capture simulator using high-pressure tubing. The compliance of the pump and tubing was checked at the maximum pressure before carrying out the planned experiments. A metal injection tube of 4 mm internal diameter and 6 mm external diameter is connected to the other end of the injection port. The presence of air bubbles in the syringe pump will affect the fracture characteristics. Hence, extreme care is taken while filling the drilling fluid into the pump and during the injection process.

### 3.3 Imaging setup

A high-resolution imaging device is used to capture the fracture evolution during the fluid injection process. The camera is placed between the four columns, with the lens facing the Perspex window. A GoPro Hero 5 camera capable of shooting fixed focus video at 120 frames per second (fps) at high resolution is used for testing the specimens. Since the camera is placed between four columns, the ambient light generates shadows and reflections on the Perspex window. Thus, an external LED light source (a white 12V LED strip producing a luminous flux of 700 lm/m) is utilized in this study to illuminate the Perspex window while also avoiding shadows and reflections. The set-up is also covered with black cloth to prevent the ambient light from entering.

## 4. Load application and instrumentation
### 4.1 Three-dimensional loading

Axial load is applied to the rigid top cap using a hydraulic actuator. The horizontal confining pressures are applied by the rubber bladders, which are filled with water to improve safety in case of leaks. A pressure multiplier, fed by an air compressor (capacity 600 kPa), also serves as an air-water interface. Pressure regulators control the differential pressure fed to the rubber bladder pairs, as shown in Fig. 1. Thus, the three stresses are controlled independently, and an anisotropic stress field is applied to the specimen.

### 4.2 Sensors and data acquisition

Two pressure transducers (2 MPa) and a load cell (100 kN) are required to measure the anisotropic boundary stresses applied onto the specimen (Fig. 1). To quantify the volume change of the specimen during the fracturing process, volumometers are used in series with the pressure multiplier, as shown in Fig. 1. A linear strain conversion transducer measures the axial displacement. The two volumometers and axial displacement transducer are used to calculate the overall volume change of the specimen during the test. The electronic pressure transducers, displacement transducers, and load cell are connected to the data acquisition system (Measurement systems – 24 bit, 16 channel, 1 kS/s), which in turn gets connected to a computer in order to record the data using LabVIEW. Another high-pressure transducer is connected between the pump and the injection port (top cap). This pressure transducer should be connected very close to the injection port to accurately measure the fracture initiation, breakdown, and propagation pressure during the injection process.

## 5. Specimen preparation
*Sand specimens*

Fraction D - Leighton Buzzard silica sand with specific gravity of 2.67, $D_{50}$ of 180 μm and uniformity coefficient of 1.38 is used to prepare the specimens. The sand had uniform gradation with sub-rounded particles. The sand specimens are artificially reconstituted using a specially designed split mould with 150 mm inner diameter and 40 mm effective height. A latex membrane of 0.2 mm thick is held around the Perspex window with o-rings and stretched over the split mould. The specimens are prepared using air pluviation (Cresswell et al. 1999, Miura and Toki 1982), and the density is controlled based on the height of fall and the rate of deposition. A constant height (10 cm) of fall with a standard 10 mm nozzle is used to deposit the sand in order to achieve repeatable loose specimens. The final void ratio achieved for this particular set of experiments is 0.88. Once the sand filled the split mould, the top cap with the injection tube is placed on the specimen. The membranes are now rolled back on the top cap and secured firmly with the help of o-rings. Care should be taken to prevent the entry of sand particles into the injection tube while locating the top cap in the sand specimen. A small cotton thread is used to plug the injection tube prior to inserting it into the sand specimen. Once a suction of 20 kPa is applied to the specimen, the split mould and the plug are removed before placing the confining chamber. An isotropic confining pressure of 20 kPa is applied to the specimen, and the applied suction is then released.

*Weak rock specimens*

The specimens are generated via microbially induced carbonate precipitation (MICP), a bio-cementation method that precipitates calcium carbonate, which acts as the binding material between the sand particles. The base material is the same sand described previously. The protocol that is followed for bio-treatment was demonstrated in Konstantinou et al. (2021). Once the bio-treated specimens are generated, they are carefully extracted from the moulds and are trimmed to eliminate potentially disturbed or uneven zones. A hole of 6 mm diameter is then drilled at the centre of the specimen with a depth of 35 mm. Glue is applied on the outer surface of the injection port in the top cap. Then, the drilled hole in the specimen and the injection tube of the top cap are aligned and left to glue together for a couple of hours before testing it.

## 6. Testing programme

*Loose sands*

The primary aim of this experimental programme is to simulate complex boundary stresses using this bespoke test rig while also capturing the fracture propagation. Hence, two distinct types of experiments are carried out using sand specimens: under cross-isotropic boundary stresses ($\sigma_a > \sigma_{r1} = \sigma_{r2}$; $\sigma_a$ - axial stress, $\sigma_{r1}$ - radial stress in direction 1, $\sigma_{r2}$ - radial stress in direction 2) and anisotropic boundary stresses ($\sigma_a > \sigma_{r1} > \sigma_{r2}$). After preparing the specimen, an isotropic boundary stress of 50 kPa is applied onto the specimen. This stage is followed by an initial injection stage (pre-fracture) which removes the particles (if any) that entered into the injection tube while inserting the tube into the specimen. Finally, the actual injection process is carried out after increasing the boundary stresses to $\sigma_a$ = 125 kPa, $\sigma_{r1}$ = $\sigma_{r2}$ = 50 kPa & $\sigma_a$ = 175 kPa, $\sigma_{r1}$ = 125 kPa, $\sigma_{r2}$ = 50 kPa for cross-isotropic and anisotropic case respectively. Independent of the anisotropic stress, the mean stress also influences the fracture characteristics. Further, wall-building injection fluid like bentonite slurry is prepared by mixing 12% of bentonite by weight of water. This fluid is injected into the sand specimen at a constant flow rate of 30 mL/min after the boundary stresses are applied.

*Weak rocks*

The intent of performing this set of experiments with weak rocks is to demonstrate the device's capability in testing materials of various properties and to study the fracture characteristics. Two tests with bio-treated sands/ weak rocks (different levels of cementation) have been carried out. The cementation level of those materials is 5% and 8.5% which has an unconfined compressive strength of 400 kPa and 1800 kPa, respectively. Both the tests are performed under the same boundary stresses (anisotropic boundary stress): $\sigma_a$ = 300 kPa, $\sigma_{r1}$ = 200 kPa and $\sigma_{r2}$ = 100 kPa. The injected fluid is solid-loaded and is formulated as follows: calcium carbonate 59 g/L, xanthan gum 1.2 g/L, magnesium oxide 0.35 g/L, water 327.4 g/L. Even though the injection fluid is different in both sand and weak rocks, the viscosity of the suspension is kept constant in each series of experiments. The injection rate is kept constant at 100 mL/min.

## 7. Results and discussion

A suite of experiments with loose sand specimens and weak rocks are performed to demonstrate the capabilities of this apparatus in simultaneously capturing the injection pressure profile and fracture morphology under different boundary stresses. The experimental study will also help understand these material systems' infiltration and fracture characteristics. Trials are carried out to ensure the repeatability of experiments. Some test parameters, such as injection hole diameter, drainage conditions, and injection fluid properties, are kept the same in all tests to eliminate any effects beyond the boundary conditions, injection rate and the materials used in this study. Pressure profiles and associated videos (at 120 fps) are captured for all the tests. Tests with loose sand specimens are carried out to understand the influence of boundary stresses on the fracture characteristics, while the tests with weak rocks are carried out to quantify the effect of material properties on the fracture behaviour.

A typical pressure profile obtained during the injection process after the application of cross-isotropic boundary stress is shown in Fig. 7. The figure shows an initial pre-injection/ pre-fracture where the injection hole is pressurized to a certain extent in order to remove the particles that might be present in the injection tube even after careful placement into the specimen. The effects of scaling borehole dimensions on fracture behaviour are ignored in the current study. Further, once the breakdown pressure is reached, the pressure held for a short period of time and then decreased. The sudden drop in the pressure post-breakdown is due to the propagation of the fracture. Following the fracture extension or instability formation, the pressure built up gradually reaching the softening/backbone curve. The continuous loop of building up pressure and sudden drop resulted from fracture propagation is also clearly seen in the accompanying images.

### 7.1 Influence of three-dimensional boundary stress

Fig. 8 and 9 show the pressure profile during fluid injection in sand and the infiltration/ cavity expansion/ fracturing regimes at different time stamps for cross-isotropic and anisotropic boundary stresses, respectively. The overall trend of the injection pressure profile between the two tests does not change. However, the slope of the softening curve is different, and the breakdown pressure is also distinct due to the different mean stress levels. The sand specimen with anisotropic boundary stress showed a higher breakdown pressure compared to cross-isotropic boundary stress due to the slightly higher mean effective stress in the former. Another important observation is that the breakdown pressure is much higher than the overall mean boundary stresses and the shear strength of the unconsolidated sand specimen. Further, the pressure does not drop below the minimum boundary stress during the fracture propagation until the test is concluded.

During the fluid injection event, since the material system used in this set of experiments is highly permeable sands, the infiltration into the medium is predominant. At the time of infiltration, the fine particles present in the injection fluid get deposited in between the pores of the granular medium. The dynamic process of particle deposition/ clogging (similar to filtration) and the material properties resulted in an increase in fracture pressure.

A cavity formation is observed when the injection pressure reached the breakdown pressure. The cavity slowly transformed into fingers/ fractures during the injection process, as shown in the images at different time stamps, which are mapped to the injection pressure. Every drop in the pressure during the injection event resulted in either increasing the size of the instability or the formation of new fingers (as shown at times $t_3$ and $t_5$ in Fig. 8). The sequence of events starts with infiltration, followed by cavity expansion, then formation of multiple fingers of which the dominating finger will propagate as fracture while the other fingers cease to grow. The fracture patterns are also distinct under cross-isotropic and anisotropic boundary stresses. In the case of cross-isotropic boundary stress, the fingers started to grow in a random direction, while in the case of anisotropic boundary stress, the fingers tend to orient in the direction of maximum radial stress (perpendicular to the minor radial stress direction as suggested by fracture models).

The infiltration and fracture area are also quantified to demonstrate the usefulness of real-time recording and tracing of the fracture. It is observed that the normalised infiltration and fracture area increases with respect to time, as shown in Fig. 10. Even though there is no noticeable change in the normalised infiltration area between the two tests, the fracture area under cross-isotropic boundary stress is larger compared to the anisotropic boundary condition. This phenomenon is primarily due to the lower mean effective stress and associated severe grain displacement. Due to the increase in fracture area, the fracture permeability increases with time. Additionally, due to particle invasion in the pores of the granular medium during the injection process, the matrix permeability gets reduced, which will result in additional formation damage in the field.

## 7.2 Influence of material properties

Two tests are performed with weak rock specimens having different cementations. The specimen with lower cementation (5%) exhibited a lower breakdown pressure, while the specimen with higher cementation (8.5%) showed a breakdown pressure about four times greater than the 5% cemented specimens, as shown in Fig. 11 and 12, respectively. The infiltration zone is observed to be smaller than in the loose sand specimens, due to the presence of cementation which reduces the permeability. Further, the initial instability is not circular, and the fracture tended to be in the direction of maximum radial stress. Due to the increase in fracture permeability in a particular direction, the infiltration front also grew in the same direction, as shown in Fig. 11. In the case of higher cementation (8.5%), during the fluid injection, multiple finger-like instabilities are formed, as shown in Fig. 12, till time ($t_5$), where it started to grow in size. However, there is a sudden increase in the length of the fracture at $t_6$ due to the increased brittleness of the material at higher amount of cementation. The associated pressure also immediately dropped very close to zero. Thus, with an increase in the cementation of weak rocks, the fracture length/ aspect ratio increased while the size of the infiltration zone decreased.

The novel fracture capture apparatus is capable of simulating the anisotropic stress conditions in the laboratory while capturing the pressure and fracture impressions simultaneously. The images that are captured during the fluid injection process will be helpful in quantitatively assessing the fracture characteristics (morphology, aspect ratio, orientation) which further can be used to calibrate some of the existing fracture models or develop new theories. The coupled hydro-mechanical behaviour of geomaterials can be captured through these novel experiments which lays foundation on the development of coupled transient models which can capture fracture propagation. Thus, the experimental apparatus will not only be useful in quantifying the fracture characteristics but also in understanding various fundamental mechanisms associated with hydraulic fracturing.

**8. Concluding remarks**

An unconventional experimental setup is designed and developed in order to enhance the fundamental understanding of hydraulic fracturing in realistic conditions. The fracture capture simulator possesses three crucial features, i.e. the application of anisotropic boundary stresses on the specimen, injecting fluid of different viscosity and flow rate while simultaneously evaluating the pressure profile and fracture characteristics. Successful implementation and satisfactory performance of the apparatus have led to the following observations and conclusions.

- Boundary stresses in the order of few MPa were successfully applied and controlled anisotropically;
- The pressure data can be directly correlated with the instability evolution due to real-time imaging;
- Multiple fingers initiate during the initial stages of fluid injection. However, with a continuous injection process, the dominating fracture progresses while the other fingers cease to grow;
- Mean effective stress, anisotropic stress and the material property plays a critical role in fracture initiation and propagation.

**Advantages of the fracture capture simulator:**

- The recording the real-time instability evolution will be helpful in quantitatively assessing the fracture characteristics such as fracture morphology, orientation, aspect ratio and fractal dimension;
- The continuous images obtained from this experimental setup can be used to identify the compaction zones around the instability by adopting cross correlation techniques. Thus, the plastic deformation boundary during fluid injection can also be identified;
- The experimental setup will help in understanding the particle migration characteristics due to different injection fluid composition. Thus, the drilling fluid can be designed to have minimum formation damage;
- Several poroelastic-plastic cavity expansion models, filtration and fracture models can be validated, or new models can be developed by accurately quantifying the infiltration and fracture regimes.

**Acknowledgements**

This work is supported and funded by bp -International Centre for Advanced Materials (bp-ICAM) on project ICAM-39 (Water injection in soft sand reservoirs) and University of Cambridge. Their support is gratefully acknowledged. We would like to thank Alistair Ross and Neil Houghton for their support and feedback in designing and fabricating this experimental setup.

**Conflict of interest**

The authors declare that they have no competing financial interests or personal relationships that could have appeared to influence the work reported in this paper.

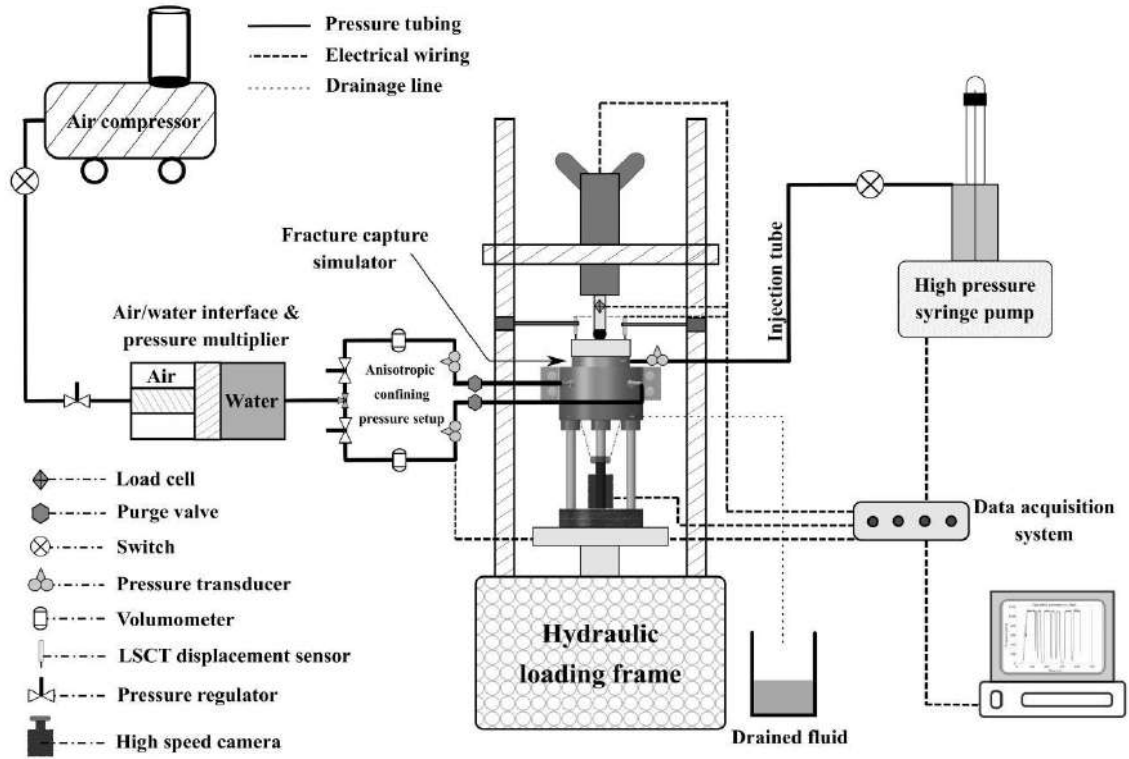

**Figure 1:** Layout of the complete fracturing set-up. The arrangement shows critical components such as fracture capture simulator, hydraulic loading frame, high pressure syringe pump, high speed/ resolution camera, air compressor, pressure multiplier, measuring instruments and sensors, and data acquisition system

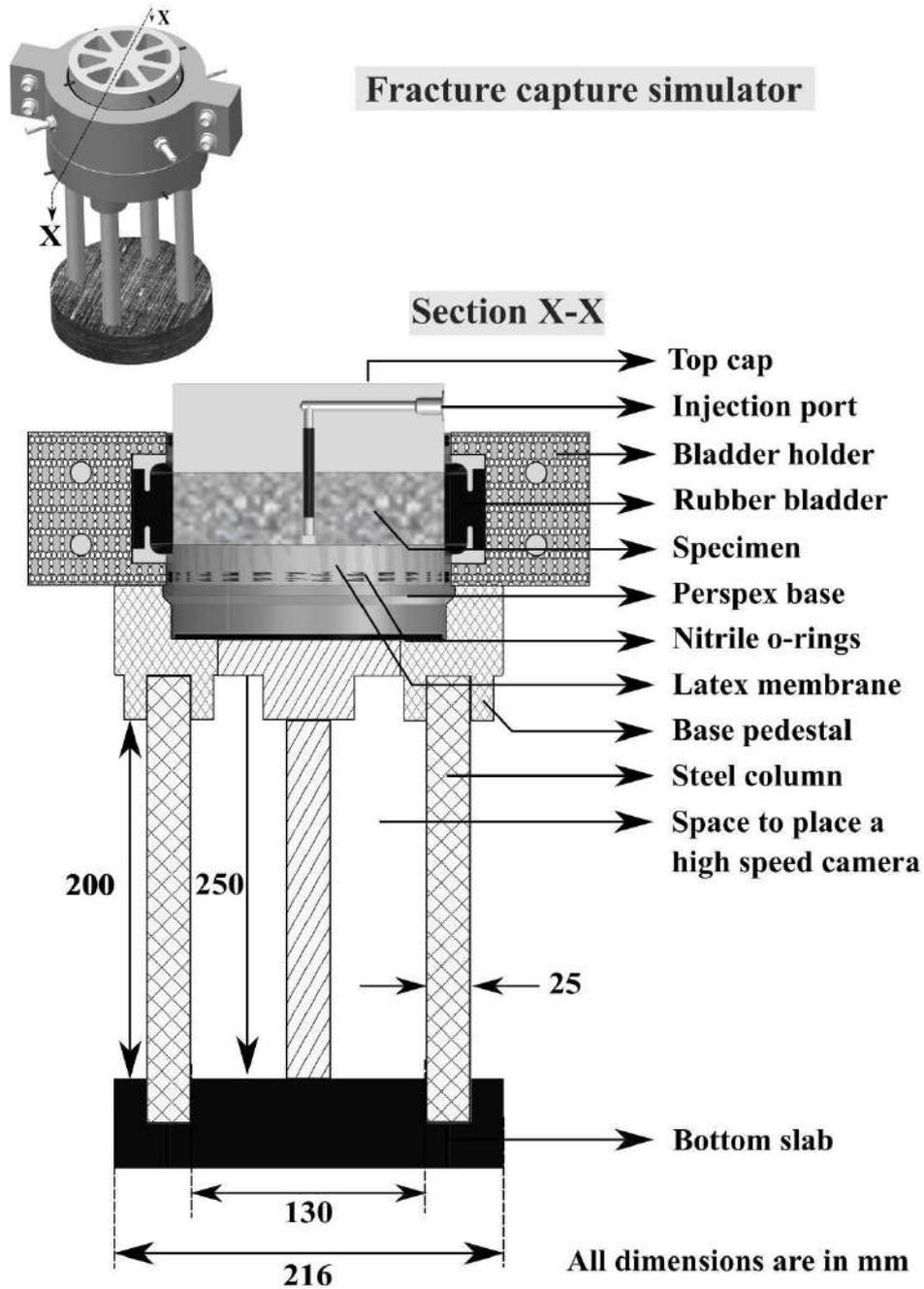

**Figure 2:** Complete sectional view of the fracture capture simulator. The set-up consists of the top cap, rubber bladders, bladder holder, base pedestal, steel columns, and a bottom slab. The space left in between the four columns is used to place the camera to capture the fracture evolution real-time

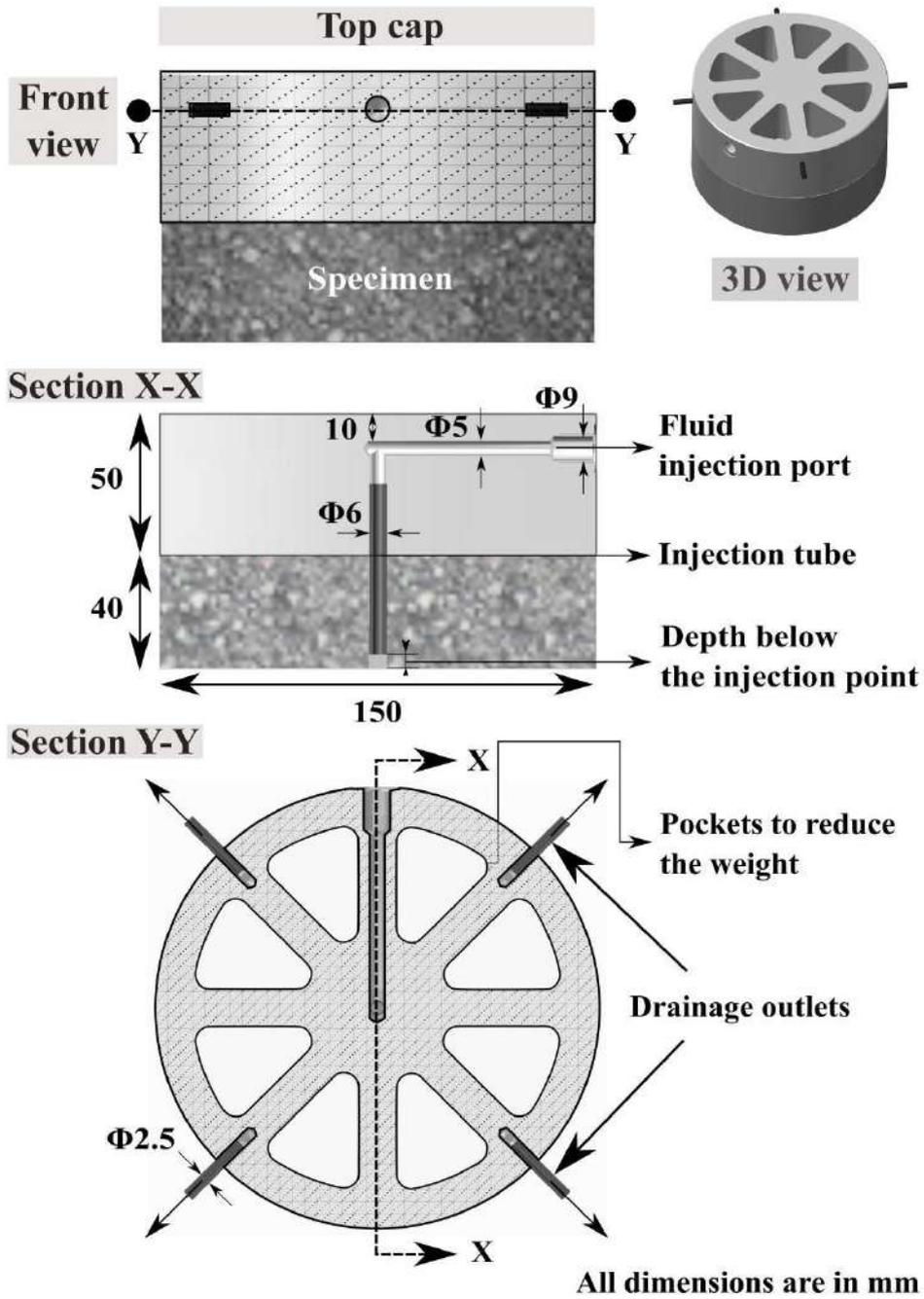

**Figure 3:** The top cap with details of the fluid injection port and drainage outlets. Cross-section XX shows the dimensions of the top cap and the injection port which takes a 90° bend to the centre of the top cap. Section YY shows the top view

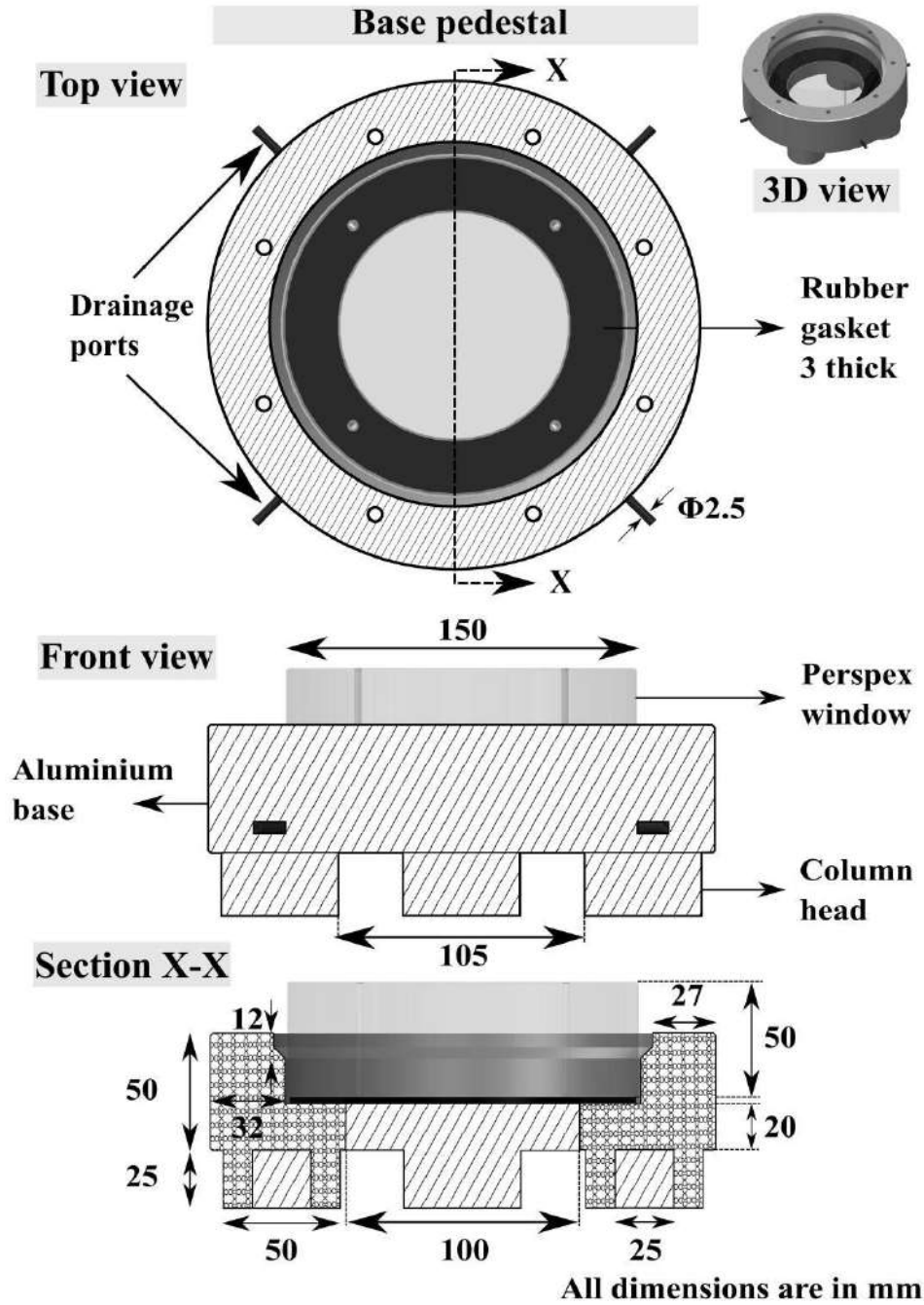

**Figure 4:** The base pedestal consists of a transparent Perspex window of 50 mm thick placed on a rubber gasket. Four drainage holes are placed 90° apart traversing the Perspex window, rubber gasket and the aluminium base. The rubber gasket is sealed with Perspex window and the aluminium base in order to prevent the leakage of the draining fluid (section XX)

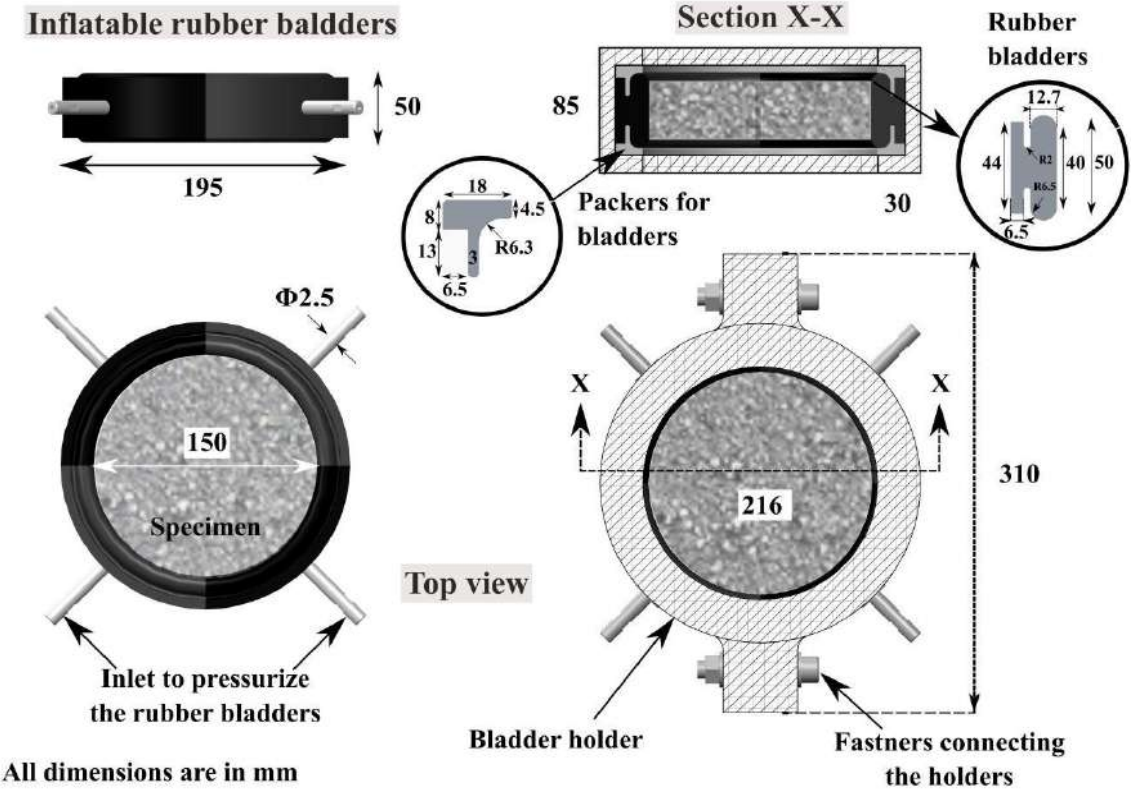

**Figure 5:** Nylon reinforced rubber bladders (4 nos.) placed around the specimen. Each rubber bladder has an inlet which gets connected to a pressure regulator, pressure multiplier and air compressor in series to apply the required pressure. The rubber bladders are held inside a bladder holder. The two halves of the bladder holder (hosting two bladders each) are connected using the fasteners. To apply uniform pressure onto the specimen, aluminium packers are placed at top and bottom to the bladders as shown in section XX. The dimensions of the packers and bladders are shown in the inset

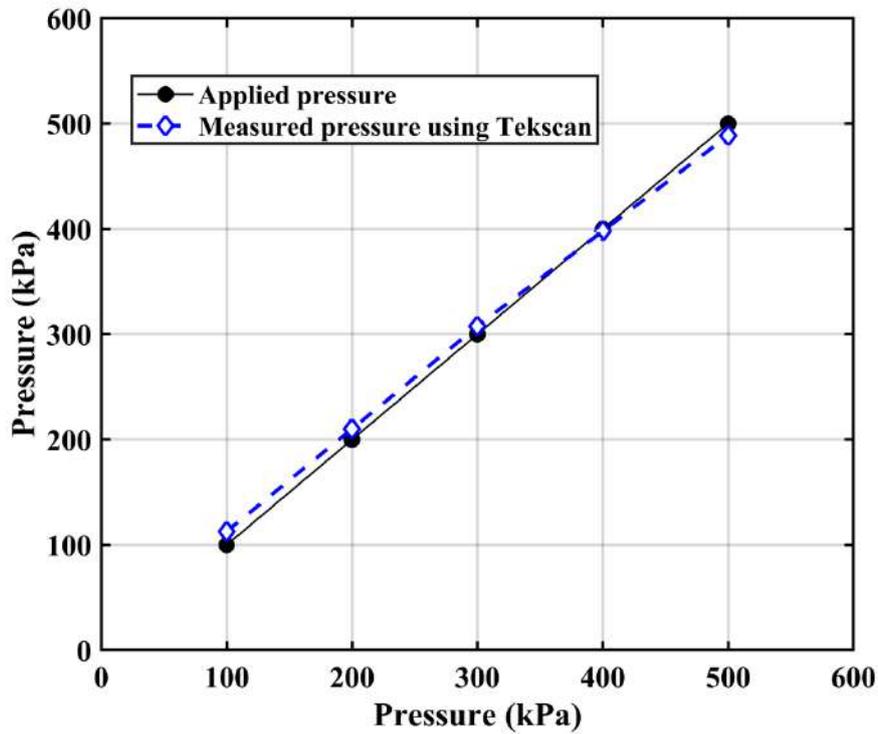

**Figure 6:** A comparison of pressure applied using the air compressor to inflate the rubber bladders and the pressure transmitted onto the specimen measured using Tekscan

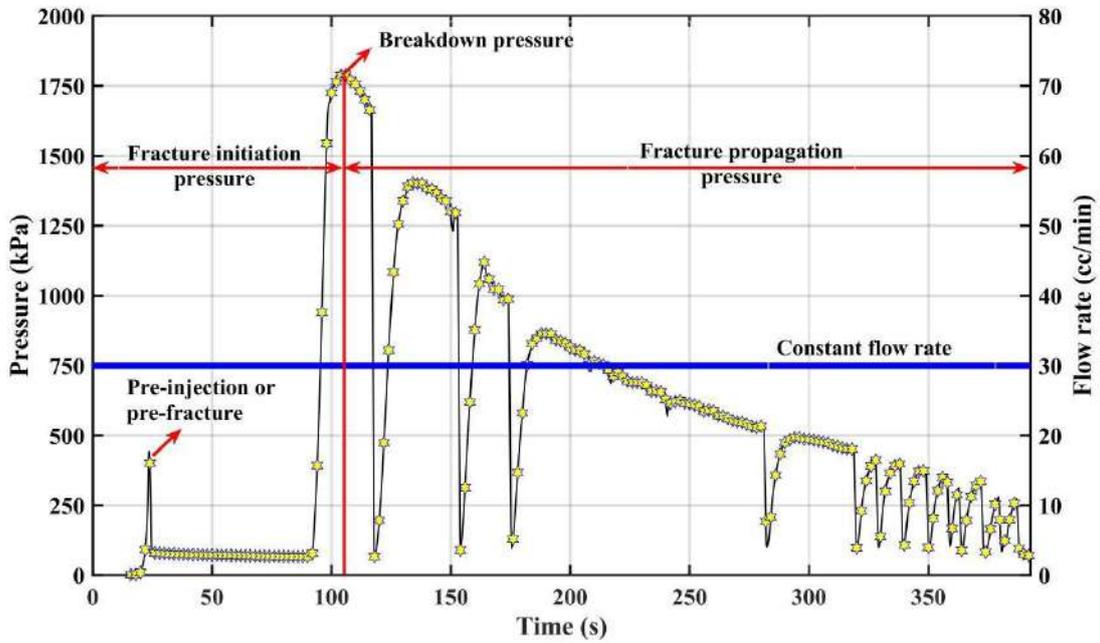

**Figure 7:** A distinct pressure profile obtained during a fracture experiment in sand after applying cross-isotropic boundary stress $\sigma_{r1} = \sigma_{r2} = 50$ kPa

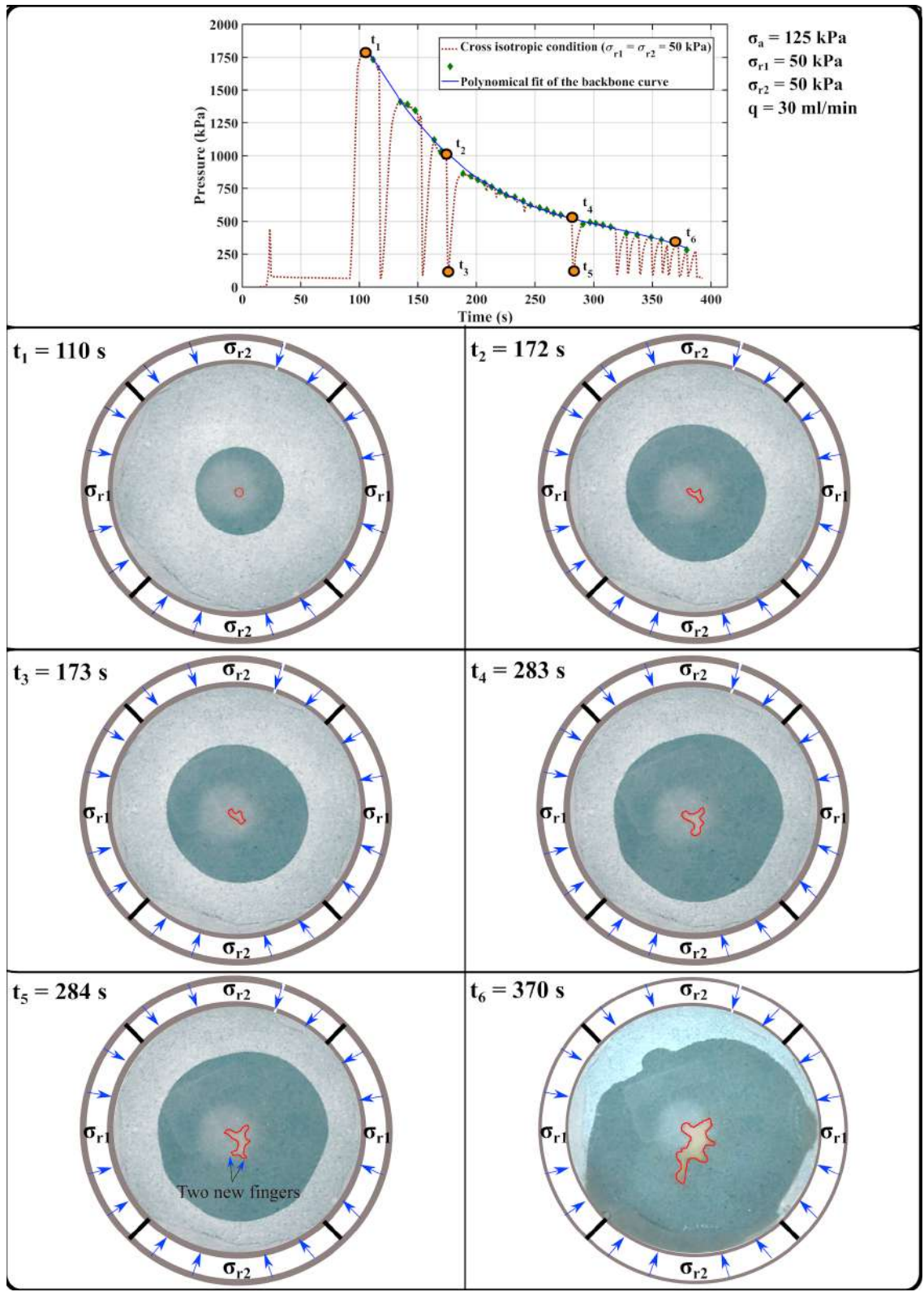

**Figure 8:** The pressure profile during the fluid injection experiment in loose sands under cross-isotropic boundary stress ($\sigma_a$ = 125 kPa, $\sigma_{r1} = \sigma_{r2}$ = 50 kPa) along with the images at different time intervals during the injection event. The growth of infiltration front and the formation of fingers especially during the pressure drop are observed

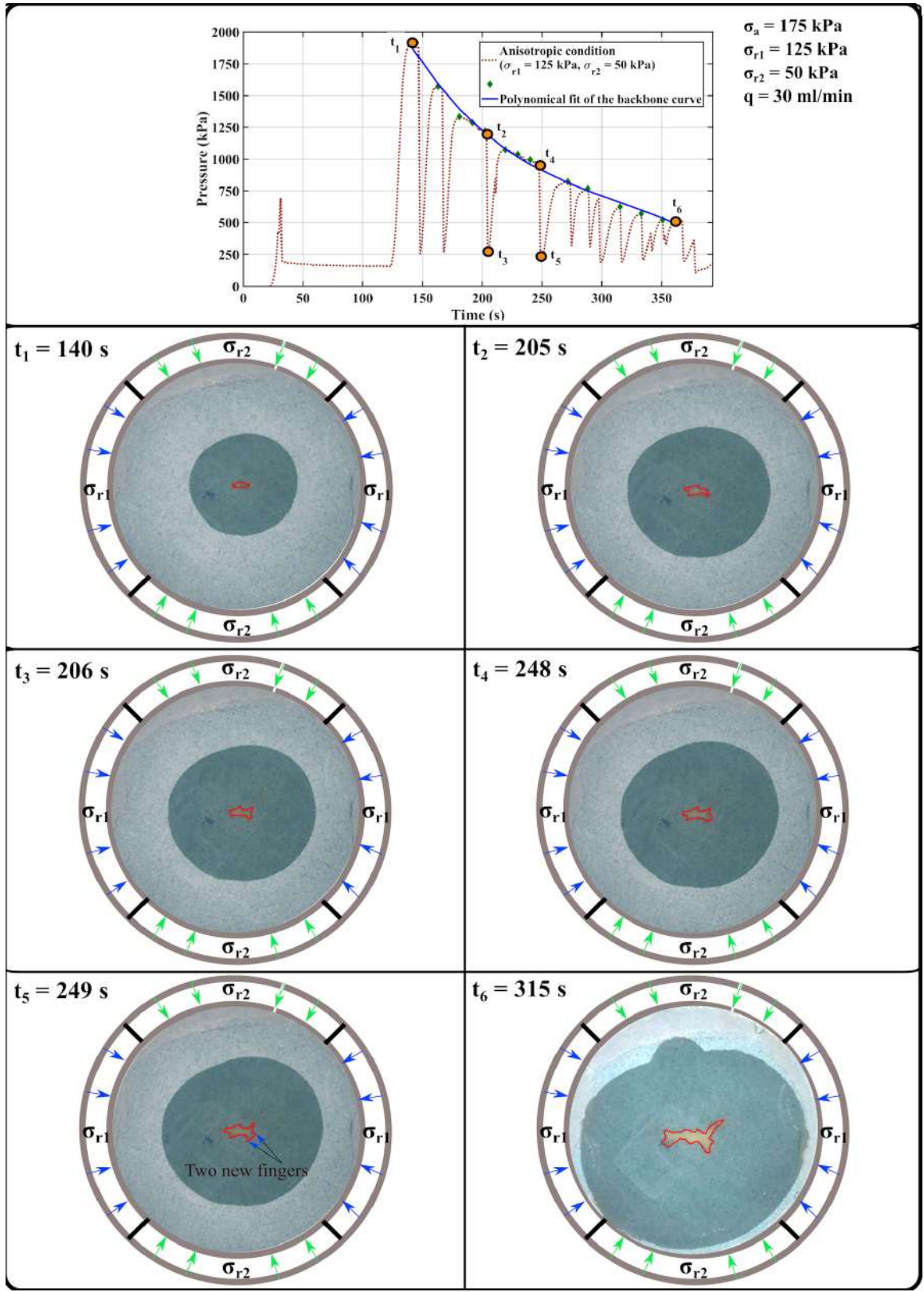

**Figure 9:** The pressure profile during the fluid injection experiment in loose sands under anisotropic boundary stress ($\sigma_a$ = 175 kPa, $\sigma_{r1}$ = 125 kPa and $\sigma_{r2}$ = 50 kPa) along with the images at different time intervals during the injection event. The growth of infiltration front and the propagation of fracture in the direction perpendicular to the minimum radial stress is observed

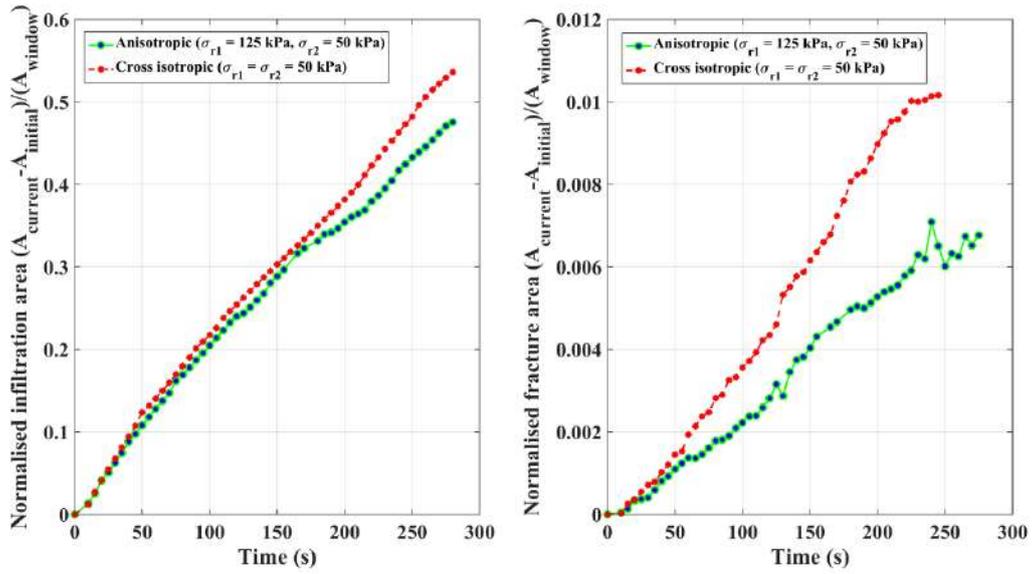

**Figure 10: Left:** Variation of normalised infiltration area with time for anisotropic and cross-isotropic boundary conditions, **Right:** Variation of normalised fracture area with time for anisotropic and cross-isotropic boundary conditions

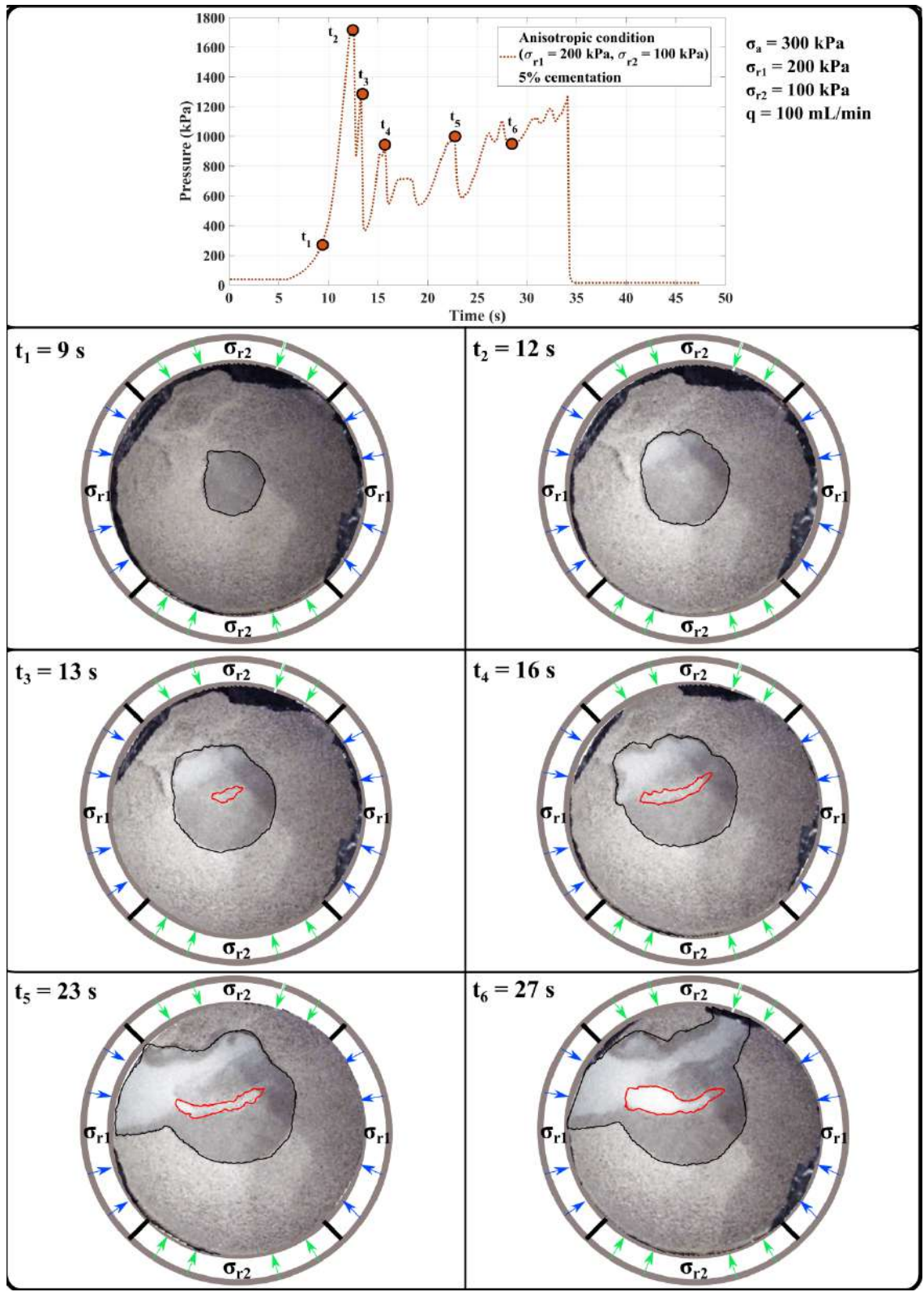

**Figure 11:** The pressure profile during the fluid injection experiment in weak rock with 5% cementation under anisotropic boundary stress ($\sigma_a$ = 300 kPa, $\sigma_{r1}$ = 200 kPa and $\sigma_{r2}$ = 100 kPa) along with the images at different time intervals during the injection event. The initiation and propagation of the fracture (marked in red) is observed along with the infiltration zone (marked in black)

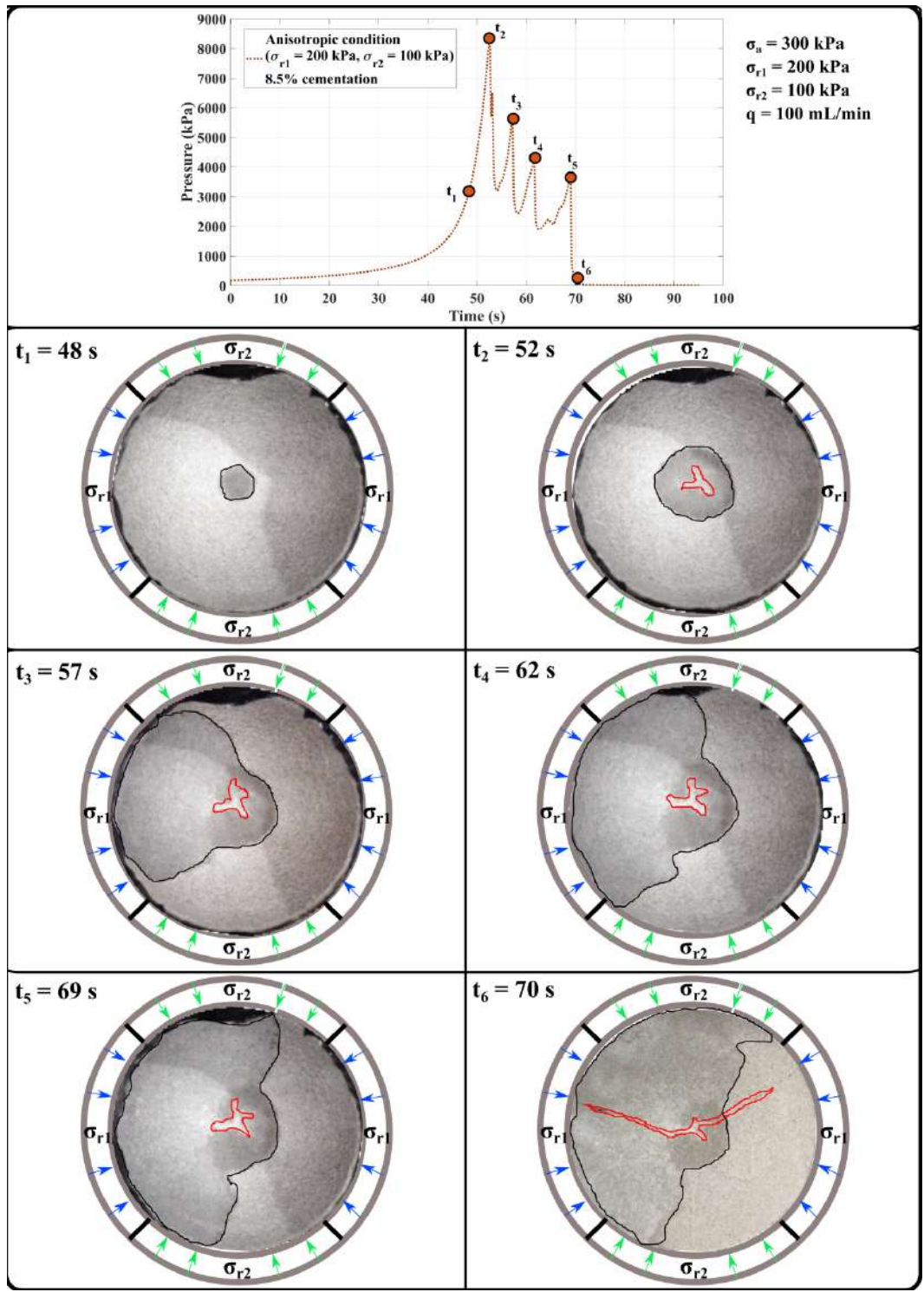

**Figure 12:** The pressure profile during the fluid injection experiment in weak rock with 8.5% cementation under anisotropic boundary stress ($\sigma_a$ = 300 kPa, $\sigma_{r1}$ = 200 kPa and $\sigma_{r2}$ = 100 kPa) along with the images at different time intervals during the injection event. The initiation and propagation of the fracture (marked in red) is observed along with the infiltration zone which grows faster in the direction of fingers